\documentclass[aps,preprintnumbers,amsmath,nofootinbib]{revtex4}
\usepackage{amssymb}
\usepackage{graphicx}
\usepackage{dcolumn}
\usepackage{bm}
\usepackage{footnpag}

\begin{document}
\title
{Azimuthally integrated HBT parameters for charged pions \\ in
nucleus-nucleus interactions versus collision energy}

\author{V.A. Okorokov} \email{VAOkorokov@mephi.ru; Okorokov@bnl.gov}
\affiliation{National Research Nuclear University "MEPhI" (Moscow
Engineering Physics Institute), Kashirskoe Shosse 31, 115409
Moscow, Russia}

\date{September 9, 2014}

\begin{abstract}
In the paper energy dependence of space-time extent of emission
region obtained from Bose\,--\,Einstein correlations is studied
for charged pions in various ion collisions for all experimentally
available energies. There is no dramatic change of HBT parameters
with increasing of collision energy per nucleon-nucleon pair,
$\sqrt{\smash[b]{s_{\footnotesize{NN}}}}$, in domain of energies
$\sqrt{\smash[b]{s_{\footnotesize{NN}}}} \geq 5$ GeV. Energy
dependence of estimations for emission duration is almost flat for
all energy domain under study within large error bars. Analytic
function is suggested for smooth approximation of energy
dependence of main HBT parameters. Fit curves demonstrate
reasonable agreement with experimental data for most HBT
parameters in energy domain
$\sqrt{\smash[b]{s_{\footnotesize{NN}}}} \geq 5$ GeV. Estimations
of some observables are obtained for energies of the LHC and FCC
project.

\textbf{PACS} 25.75.-q,
25.75.Gz,
25.75.Nq
\end{abstract}

\maketitle

\section{Introduction}\label{intro}
At present femtoscopic correlations in particular that based on
Bose\,--\,Einstein correlations are unique experimental method for
the determination of sizes and lifetimes of sources in high energy
and nuclear physics. The discussion below is focused on specific
case of femtoscopy, namely, on correlations in pairs of identical
charged pions with small relative momenta -- HBT-interferometry --
in nucleus-nucleus collisions. Space-time characteristics for
emission region of secondary particles created in (heavy) ion
collisions are important for study of deconfinement state of
strongly interacting matter -- strong-coupling quark-gluon plasma
(sQGP). Furthermore the study of energy dependence of femtoscopic
observables can be useful for understanding in detail the
transition from sQGP produced at higher energies to confined
hadronic resonance matter created in final state at lower
energies. HBT analysis allows studying dynamic features of
interaction process at late, i.e. soft, stage of space-time
evolution of multiparticle final state. Therefore the study of
nucleus-nucleus collisions in wide energy domain by correlation
femtoscopy seems important for better understanding both the
equation of state (EOS) of strongly interacting matter and general
dynamic features of soft processes.

The paper is organized as follows. In Sec.\,II, definitions of
main observables for correlation femtoscopy are briefly described.
The Sec.\,III devotes discussion of experimental energy dependence
for space-time extent of source of charged pions and corresponding
fits. Also estimations for femtoscopic observables are shown for
the LHC and the Future Circular Collider (FCC) project energies.
Some final remarks and conclusions are presented in Sec.\,IV.

\section{Method and variables}\label{sec:2}
In general phenomenological
parameterization of correlation function (CF) for two identical particles with 4-momenta $p_{\,1}, p_{\,2}$ and with taking into account different forms of
corrections on Coulomb final state interaction (FSI) can be written as follows
\cite{Okorokov-arXiv-1312.4269}:
\begin{equation}
C_{2,(m)}^{\mbox{\scriptsize{ph}}}(q,K)=\epsilon
P_{\mbox{\footnotesize{coul}}}^{(m)}(q)
\bigl[\epsilon^{-1}+\mathbf{K}_{2}^{\mbox{\scriptsize{ph}}}({\bf
A})\bigr], ~~\epsilon=\left\{
\begin{array}{ll}
\lambda, & \mbox{at}~m=1,2;\\
1, & \mbox{at}~m=3.
\end{array}
\right.\label{eq:2.7}
\end{equation}
where $m=1$ corresponds to the standard Coulomb correction, $m=2$
-- the dilution procedure and $m=3$ -- the Bowler--Sinyukov
correction, $q \equiv (q^{0},\vec{q})=p_{\,1}-p_{\,2}$ is the
relative 4-momentum, $K \equiv
(K^{0},\vec{K})=(p_{\,1}+p_{\,2})/2$ -- the average 4-momentum of
particles in pair (pair 4-momentum), for standard simplest
(Gaussian) case
\begin{equation}
\mathbf{K}_{2}^{\mbox{\scriptsize{ph}}}({\bf
A})=\prod\limits_{i,j=1}^{3}
\mathbf{K}_{2}^{\mbox{\scriptsize{ph}}}(A_{ij})=
\exp\biggl(-\sum\limits_{i,j=1}^{3}q_{i}R_{ij}^{\,2}q_{j}\biggr).
\label{eq:2.3}
\end{equation}
Here ${\bf A} \equiv \vec{q}\,{\bf R}^{2}\vec{q}^{\,T}$ and ${\bf
R}^{2}$ are the matrices $3 \times 3$, $\vec{q}^{\,T}$ --
transposed vector $\vec{q}$, $\forall~ i,j:
R^{\,2}_{ij}=R^{\,2}_{ji}, R^{\,2}_{ii} \equiv R^{\,2}_{i}$, where
$R_{i}=R_{i}(K)$ are parameters characterized the linear scales of
homogeneity region \cite{Sinyukov-NATOSeries-346-309-1995}; the
products are taken on space components of vectors,
$\lambda(K)=\mathbf{K}_{2}(0,K), 0 \leq \lambda \leq 1$ is the
parameter which characterizes the degree of source chaoticity.
Taking into account the hypothesis of cylindrical symmetry of
source the volume of homogeneity region is derived as follows
\cite{Okorokov-ISHEPP-101-2006}
\begin{equation}
V=(2\pi)^{3/2}\prod\limits_{i=1}^{3}R_{i}. \label{eq:2.4}
\end{equation}

The space component of pair 4-momentum ($\vec{K}$) is decomposed
on longitudinal
$k_{\parallel}=(p_{\,\parallel,1}+p_{\,\parallel,2})/2$ and
transverse
$\vec{k}_{\perp}=(\vec{p}_{\perp,1}+\vec{p}_{\perp,2})/2$ parts of
pair momentum. In the paper the decomposition of
Pratt\,--\,Bertsch
\cite{Pratt-PRD-33-1314-1986,Bertsch-PRC-37-1896-1988} is used for
$\vec{q}$ as well as the longitudinal co-moving system (LCMS)
frame. The radius $R_{\mbox{\scriptsize{o}}}$ contains additional
contribution from the temporal extent of the source. Therefore
this parameter is usually excluded from calculation of $V$ and the
volume of source can be written as follows
\begin{equation}
V=(2\pi)^{3/2}R_{\mbox{\scriptsize{s}}}^{2}R_{\mbox{\scriptsize{l}}}.
\label{eq:2.4new}
\end{equation}
As seen
$V^{(\ref{eq:2.4})}/V^{(\ref{eq:2.4new})}=R_{\mbox{\scriptsize{o}}}/R_{\mbox{\scriptsize{s}}}$,
where $V^{(\ref{eq:2.4})}$ denotes the source volume calculated
from eq.\,(\ref{eq:2.4}) and the volume of emission region defined
in accordance with eq.\,(\ref{eq:2.4new}) is designated by
$V^{(\ref{eq:2.4new})}$. But it should be emphasized that in the
limit for absolute value of transverse pair momentum vector
$k_{\perp} \to 0$, no transverse vector allows to distinguish
between out- and side-components
\cite{Wiedemann-PR-319-145-1999,Okorokov-UchPosob-2009}. This
implies that $\lim_{k_{\perp} \to 0}
R_{\mbox{\scriptsize{o}}}(K)=\lim_{k_{\perp} \to 0}
R_{\mbox{\scriptsize{s}}}(K)$. Consequently, it is expected
$V^{(\ref{eq:2.4})} \approx V^{(\ref{eq:2.4new})}$ for particles
with low $k_{\perp}$ and both the relations (\ref{eq:2.4}) and
(\ref{eq:2.4new}) for freeze-out volume are valid for such
particles.

The one of the important additional observables is the following
difference \cite{Wiedemann-PR-319-145-1999,Okorokov-UchPosob-2009}
\begin{equation}
\delta \equiv
R_{\mbox{\scriptsize{o}}}^{2}-R_{\mbox{\scriptsize{s}}}^{2}.
\label{eq:2.5dop}
\end{equation}
If the emission function features no position-momentum
correlation, then $\delta$ is finite at non-zero $\vec{K}$ only
due to explicit $\vec{K}$-dependence (resulting from the
mass-shell constraint $q^{0}=\vec{q}\vec{K}/K^{0}$)
\cite{Wiedemann-PR-319-145-1999}. In this case
\begin{equation}
\delta \approx \beta_{\perp}^{2}(\Delta\tau)^{2},
\label{eq:2.6dop}
\end{equation}
where $\beta_{\perp}=k_{\perp}/m_{\perp}$ is the transverse
velocity of pair of particles with mass $m$,
$m_{\perp}^{2}=k_{\perp}^{2}+m^{2}$, $\Delta\tau$ -- the emission
duration for the particle type under discussion. It should be
stressed the last relation is valid in some specific cases 1D
hydrodynamics while is violated in both the cascade approaches and
multidimensional hydrodynamic models. Thus in the framework of
some assumptions the $\delta$ gives direct access to the emission
duration of the source and allows to partially disentangle the
spatial and temporal information contained in radii parameters
$R_{ij}$ \cite{Wiedemann-PR-319-145-1999}. The sensitivity to the
$\Delta\tau$ is the main advantage of the observable
(\ref{eq:2.5dop}).

In the paper the following set of main femtoscopic observables
$\mathcal{G}_{1} \equiv
\{\mathcal{G}_{1}^{i}\}_{i=1}^{4}=\{\lambda,
R_{\mbox{\scriptsize{s}}}, R_{\mbox{\scriptsize{o}}},
R_{\mbox{\scriptsize{l}}}\}$ is under consideration as well as the
set of some important additional observables which can be
calculated with help of HBT radii $\mathcal{G}_{2} \equiv
\{\mathcal{G}_{2}^{j}\}_{j=1}^{3}=\{R_{\mbox{\scriptsize{o}}}/
R_{\mbox{\scriptsize{s}}}, \Delta\tau, V\}$. The set of parameters
$\mathcal{G}_{1}$ characterizes the chaoticity of source and its
4-dimensional geometry at freeze-out stage completely. Scaled
femtoscopic parameters $\mathcal{G}_{1}^{i}$, $i=2-4$, $\delta$
and $\mathcal{G}_{2}^{3}$ are calculated as follows
\cite{Okorokov-arXiv-1312.4269}:
\begin{equation}
R_{i}^{n}=R_{i}/R_{\mbox{\scriptsize{A}}},~
i=\mbox{s,o,l};~~~\delta^{n}=\delta/R_{\mbox{\scriptsize{A}}}^{2};~~~V^{n}=V/V_{\mbox{\scriptsize{A}}}.\label{eq:2.8}
\end{equation}
Here $R_{\mbox{\scriptsize{A}}}=r_{0}A^{1/3},
V_{\mbox{\scriptsize{A}}}=4\pi R^{3}_{\mbox{\scriptsize{A}}}/3$ is
radius and volume of spherically-symmetric nucleus, $r_{0}=(1.25
\pm 0.05)$ fm \cite{Valentin-book-1982,Mukhin-book-1983}. The
change $R_{\mbox{\scriptsize{A}}} \to \langle
R_{\mbox{\scriptsize{A}}}\rangle=0.5(R_{\mbox{\scriptsize{A}}_{1}}+R_{\mbox{\scriptsize{A}}_{2}})$
is made in the relation (\ref{eq:2.8}) in the case of
non-symmetric nucleus-nucleus collisions
\cite{Okorokov-arXiv-1312.4269}. One needs to emphasize the most
central collisions are usually used for study the space-time
characteristics of final-state matter, in particular, for
discussion of global energy dependence of femtoscopic observables
(see below Sec.\,3). Thus the using of radius of all the nucleus
in (\ref{eq:2.8}) seems reasonable. In general case the scale
factor in (\ref{eq:2.8}) for calculation of normalized femtoscopic
radii, $\delta$ and volume should takes into account the
centrality of nucleus-nucleus collisions. The normalization
procedure suggested in \cite{Okorokov-arXiv-1312.4269} allows to
consider two data samples, namely, (i) only (quasi)symmetric heavy
ion collisions and (ii) all available data for nucleus-nucleus
collisions. Ensemble of experimental data reviewed in
\cite{Okorokov-arXiv-1312.4269} with replacement of $\mbox{Au+Au}$
points at $\sqrt{\smash[b]{s_{\footnotesize{NN}}}} \geq 11.5$ GeV
by the recent STAR results of high-statistics analysis for
$\mbox{Au+Au}$ collisions at
$\sqrt{\smash[b]{s_{\footnotesize{NN}}}}=7.7, 11.5-62.4$ and 200
GeV \cite{STAR-arXiv-1403.4972} is used in the present study.
%
%
It should be noted that estimation based on
\cite{Aamodt-PLB-696-328-2011} is used for $\lambda$ at
$\sqrt{\smash[b]{s_{\footnotesize{NN}}}}=2.76$ TeV here as well as
in the previous analysis \cite{Okorokov-arXiv-1312.4269}. But
recent study at the LHC energy \cite{Abelev-PRC-89-024911-2014}
obtains slightly larger value of chaoticity ($\lambda \sim 0.6$)
for two-pion correlations in kinematic domain under study. The
change of the $\lambda$ value in TeV-region of collision energies
can influence on general trend and quantitative results discussed
below for this femtoscopic parameter. Therefore this discrepancy
should be investigated additionally. Such study is in the
progress.

\section{Energy dependence of space-time extent of pion source}\label{sec:3}
\begin{figure*}
\includegraphics[width=15.5cm,height=17.0cm]{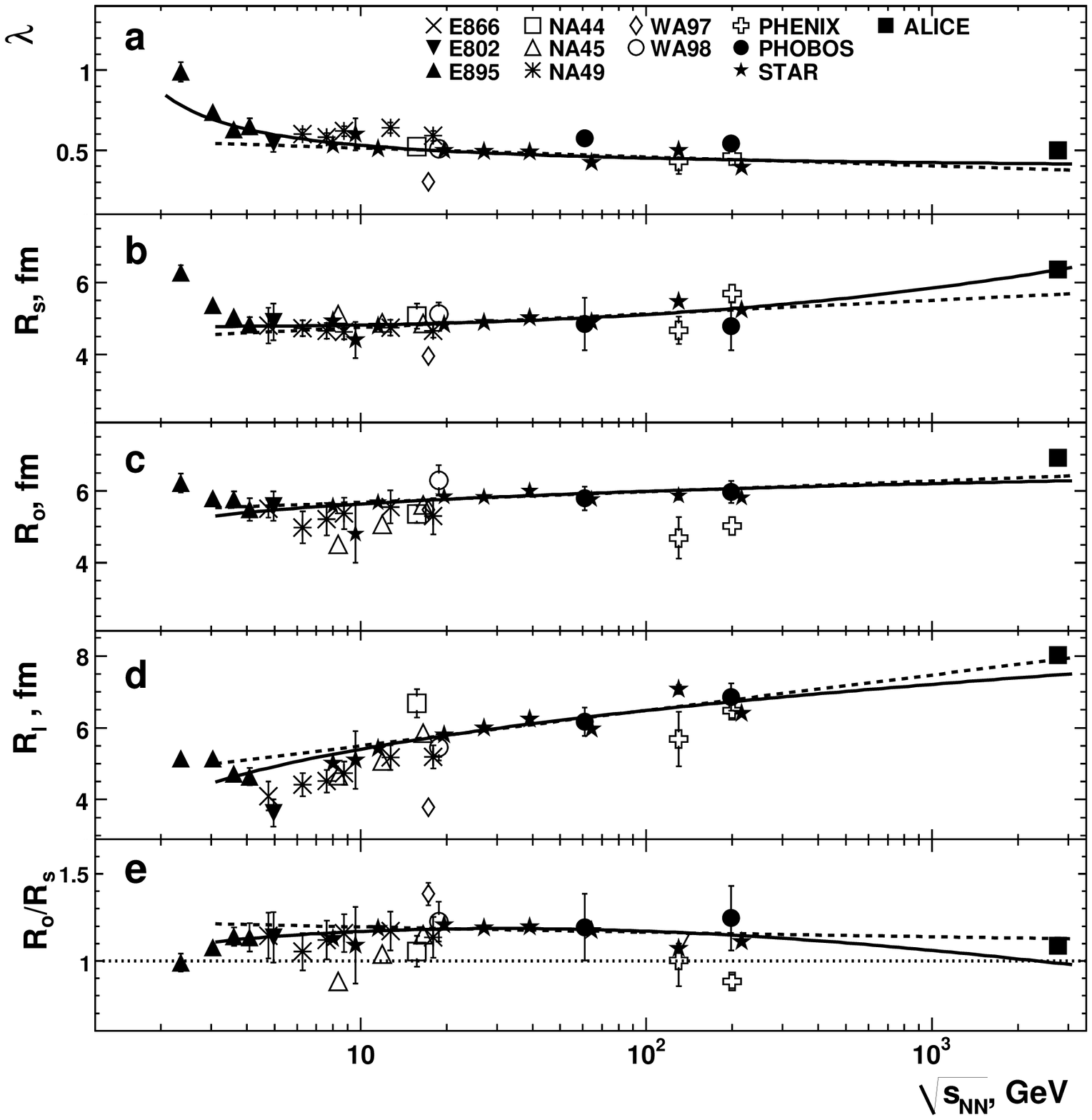}
\vspace*{8pt} \caption{Dependence of chaoticity parameter (a),
HBT-radii (b -- d) and ratio
$R_{\mbox{\scriptsize{o}}}/R_{\mbox{\scriptsize{s}}}$ (e) on
collision energy for central heavy ion $\mbox{Au+Au, Au+Pb,
Pb+Pb}$ interactions at midrapidity and $\langle k_{\perp}\rangle
\simeq 0.2$ GeV/$c$
\cite{Okorokov-arXiv-1312.4269,STAR-arXiv-1403.4972}. Experimental
results are demonstrated for pairs of $\pi^{-}$ mesons (in the
cases of ALICE, STAR at
$\sqrt{\smash[b]{s_{\footnotesize{NN}}}}=7.7-62.4$ and 200 GeV --
for $\pi^{\pm}\pi^{\pm}$ pairs) and for standard Coulomb
correction $P_{\mbox{\scriptsize{C}}}^{(1)}(q)$ (in the cases of
ALICE, NA44, NA45, PHOBOS and STAR at
$\sqrt{\smash[b]{s_{\footnotesize{NN}}}}=7.7, 11.5-62.4$ and 200
GeV -- for correction $P_{\mbox{\scriptsize{C}}}^{(3)}$).
Statistical errors are shown (for NA44 -- total uncertainties).
The solid lines (a -- d) correspond to the fits by function
(\ref{eq:Fit-1}) and dashed lines -- to the fits by specific case
$\mathcal{G}_{1}^{i} \propto \ln\varepsilon$, $i=1 - 4$. Fitted
data samples for $\lambda$ (a) and for $R_{\mbox{\scriptsize{l}}}$
(d) do not include the point of the WA97 experiment
\cite{Antinori-JPGNPP-27-2325-2001} while the fits for transverse
HBT radii (b, c) are shown for samples with point from
\cite{Antinori-JPGNPP-27-2325-2001}. Smooth solid and dashed
curves at (e) correspond to the ratio
$R_{\mbox{\scriptsize{o}}}/R_{\mbox{\scriptsize{s}}}$ calculated
from the fit results for $R_{\mbox{\scriptsize{s}}}$ and
$R_{\mbox{\scriptsize{o}}}$, dotted line is the level
$R_{\mbox{\scriptsize{o}}}/R_{\mbox{\scriptsize{s}}}=1$.}
\label{fig:1}
\end{figure*}

Dependencies of femtoscopic parameters
$\mathcal{G}_{1}^{i}(\sqrt{\smash[b]{s_{\footnotesize{NN}}}})$,
$i=1-4$ and
$R_{\mbox{\scriptsize{o}}}/R_{\mbox{\scriptsize{s}}}(\sqrt{\smash[b]{s_{\footnotesize{NN}}}})$
are shown in Figs.\,\ref{fig:1}a -- d and Fig.\,\ref{fig:1}e
respectively. It should be stressed that the STAR results
considered here were obtained for fit function (\ref{eq:2.3}) with
taking into account two additional cross-terms for
$R_{\mbox{\scriptsize{os}}}$ and $R_{\mbox{\scriptsize{ol}}}$ as
well as with improved Coulomb correction
$P_{\mbox{\scriptsize{coul}}}^{(3)}(q)$. The two sets of STAR
results for $\{\mathcal{G}_{1}^{i}\}_{i=1}^{4}$ are in a good
agreement for previously study \cite{Okorokov-arXiv-1312.4269} and
for present analysis for most femtoscopic parameters under
consideration in Fig.\,\ref{fig:1}. The some decreasing is seen
for $\lambda$ and transverse radii $R_{\mbox{\scriptsize{o}}},
R_{\mbox{\scriptsize{s}}}$ for $\mbox{Au+Au}$ collisions at
$\sqrt{\smash[b]{s_{\footnotesize{NN}}}}=200$ GeV with respect to
the earlier STAR results from \cite{Okorokov-PRC-71-044906-2005}.
The new results from \cite{STAR-arXiv-1403.4972} agree better with
both the general trends and the results of other experiments
(PHENIX and PHOBOS) at top RHIC energy
$\sqrt{\smash[b]{s_{\footnotesize{NN}}}}=200$ GeV. The chaoticity
parameter $\lambda$ decreases with increasing
$\sqrt{\smash[b]{s_{\footnotesize{NN}}}}$ relatively fast at lower
(AGS) energies and shows the weak changing at
$\sqrt{\smash[b]{s_{\footnotesize{NN}}}} > 4$ GeV
(Fig.\,\ref{fig:1}a). Femtoscopic radii of source in transverse
plane with respect to the beam direction,
$R_{\mbox{\scriptsize{s}}}$ (Fig.\,\ref{fig:1}b) and
$R_{\mbox{\scriptsize{o}}}$ (Fig.\,\ref{fig:1}c), show little
change over a wide range of energy $5 \lesssim
\sqrt{\smash[b]{s_{\footnotesize{NN}}}} \lesssim 200$ GeV which
corresponds to highest AGS -- SPS -- RHIC beam collision energies.
On the other hand, the value of source size in longitudinal
direction, $R_{\mbox{\scriptsize{l}}}$ (Fig.\,\ref{fig:1}d),
appears to reach a minimum around
$\sqrt{\smash[b]{s_{\footnotesize{NN}}}}=5$ GeV, rising in energy
domain available at RHIC. As seen there is increasing of HBT radii
(Figs.\,\ref{fig:1}b -- d) at growth of collision energy from
$\sqrt{\smash[b]{s_{\footnotesize{NN}}}} \sim 20$ GeV up to
maximum available LHC energy
$\sqrt{\smash[b]{s_{\footnotesize{NN}}}}=2.76$ TeV. The
significant increasing of HBT radii is seen for much broader
energy range (on about two order of magnitude
$\sqrt{\smash[b]{s_{\footnotesize{NN}}}} \sim 0.02 - 3$ TeV) only
than it was expected early at the beginning of RHIC operation.
Therefore the space-time extent of emission region at freeze-out
changes slowly at increasing of collision energy. The transverse
radius $R_{\mbox{\scriptsize{s}}}$ reflects the spatial extent of
particle source, whereas $R_{\mbox{\scriptsize{o}}}$ is also
affected by dynamics
\cite{Retiere-PRC-70-044907-2004,Mount-PRC-84-014908-2011} and is
believed to be related to the duration of particle emission
\cite{Bertsch-NPA-498-173c-1989}. As indicated, for example, in
\cite{STAR-arXiv-1403.4972}, the ratio
$R_{\mbox{\scriptsize{o}}}/R_{\mbox{\scriptsize{s}}}$ was
predicted to increase with beam energy by hydrodynamical
calculations and might shows an significant enhancement if the
life-time of the collision evolution (and, within these models,
the duration of particle emission as a result) was to be extended
by entrance into a different phase
\cite{Bertsch-NPA-498-173c-1989}. There is no significant
increasing of ratio
$R_{\mbox{\scriptsize{o}}}/R_{\mbox{\scriptsize{s}}}$ in all
experimentally available energy domain (Fig.\,\ref{fig:1}e).
Recent developments, in particular in viscous hydrodynamics, allow
to get reasonable agreement between experimental and model values
of $R_{\mbox{\scriptsize{o}}}/R_{\mbox{\scriptsize{s}}}$ at top
RHIC energy and demonstrate that the behavior of experimental
dependencies of
$R_{\mbox{\scriptsize{o}}}/R_{\mbox{\scriptsize{s}}}$ on kinematic
variables can be explained in particular by realistic EoS with
crossover phase transition and sQGP at high temperatures
\cite{Broniowski-PRL-101-022301-2008,Pratt-PRL-102-232301-2009,Karpenko-PLB-688-50-2010,
Karpenko-PRC-81-054903-2010,Werner-PRC-82-044904-2010,Bozek-PRC-83-044910-2011}.
Therefore the soft femtoscopic observables confirm the phase
transition and creation of deconfinement state of strongly
interacting matter in collider experiments.

\begin{figure*}
\includegraphics[width=15.5cm,height=17.0cm]{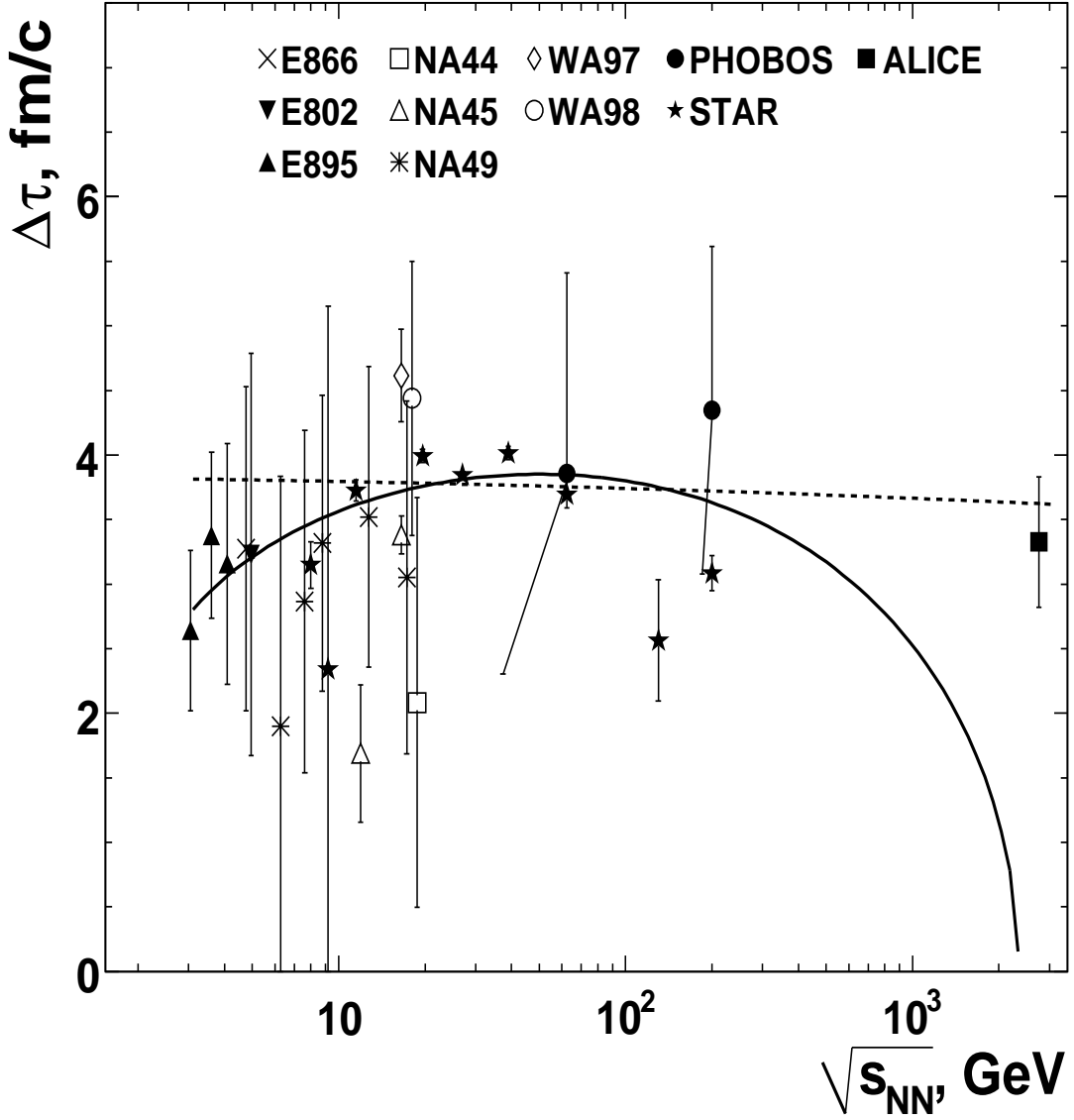}
\vspace*{8pt} \caption{Energy dependence of emission duration for
secondary charged pions in central heavy ion collisions
$\mbox{Au+Au, Au+Pb, Pb+Pb}$ in midrapidity region and at $\langle
k_{\perp}\rangle \simeq 0.2$ GeV/$c$. Experimental results are
shown for the same particle types and Coulomb corrections as well
as in Fig.\,\ref{fig:1}. Error bars are only statistical (for NA44
-- total uncertainties). Smooth curves are derived from
(\ref{eq:2.6dop}) and the fit results for
$R_{\mbox{\scriptsize{s}}}$, $R_{\mbox{\scriptsize{o}}}$ without
the point of the WA97 experiment
\cite{Antinori-JPGNPP-27-2325-2001}. The solid line corresponds to
the fits of HBT radii by function (\ref{eq:Fit-1}) and dashed line
-- to the fits by specific case $R_{i} \propto \ln\varepsilon$,
$i=\mbox{s}, \mbox{o}$.} \label{fig:2}
\end{figure*}

Taking into account the view of experimental dependencies in
Figs.\,\ref{fig:1}a -- d the following function is suggested
\begin{equation}
f(\sqrt{\smash[b]{s_{\footnotesize{NN}}}}) = a_{1}\left[1 +
a_{2}(\ln\varepsilon)^{a_{3}}\right] \label{eq:Fit-1}
\end{equation}
for smooth approximation of
$\mathcal{G}_{1}^{i}(\sqrt{\smash[b]{s_{\footnotesize{NN}}}})$,
$i=1-4$, where $\varepsilon \equiv s_{\footnotesize{NN}}/s_{0}$,
$s_{0}=1$ GeV$^{2}$. Also the specific case of (\ref{eq:Fit-1}) at
$a_{3}=1.0$ is under consideration. As seen from
Figs.\,\ref{fig:1}b -- d there is indication on change of behavior
of energy dependence (inflection point) for
$\{\mathcal{G}_{1}^{i}\}_{i=2}^{4}$ at
$\sqrt{\smash[b]{s_{\footnotesize{NN}}}} \simeq 5$ GeV. This
inflection point is seen most clear for
$R_{\mbox{\scriptsize{l}}}$ (Fig.\,\ref{fig:1}d). Therefore the
fit function (\ref{eq:Fit-1}) is used for approximation of the
energy dependence of HBT radii in the energy domain
$\sqrt{\smash[b]{s_{\footnotesize{NN}}}} \geq 5$ GeV only.
Experimental energy dependence of $\lambda$ is fitted by general
function (\ref{eq:Fit-1}) at all available energies. As seen the
point from the WA97 experiment \cite{Antinori-JPGNPP-27-2325-2001}
differs significantly from other results at close energies for
$\lambda$ (Fig.\,\ref{fig:1}a) and longitudinal radius
(Fig.\,\ref{fig:1}d). Thus for these parameters fits are made for
data sample (i) with exception of the point from
\cite{Antinori-JPGNPP-27-2325-2001}. For each main HBT parameters
$\{\mathcal{G}_{1}^{i}\}_{i=1}^{4}$ fits are made for both the
statistical and total errors, where total errors of experimental
points include available clear indicated systematic errors added
in quadrature to statistical ones. The numerical values of fit
parameters are presented in Table\,\ref{tab:1}, where the second
line for each HBT parameter $\{\mathcal{G}_{1}^{i}\}_{i=1}^{4}$
corresponds to the approximation by specific case of
(\ref{eq:Fit-1}). Fit curves are shown in Fig.\,\ref{fig:1} by
solid lines for (\ref{eq:Fit-1}) and by dashed lines for specific
case of fit function at $a_{3}=1.0$ with taking into account the
statistical errors. In general fit function described above agrees
reasonably with experimental dependence
$\mathcal{G}_{1}^{i}(\sqrt{\smash[b]{s_{\footnotesize{NN}}}})$,
$i=1-4$ (Fig.\,\ref{fig:1}a -- d). But the fit qualities are poor
for all the main HBT parameters, especially for $\lambda$, with
statistical errors taken into account (Table\,\ref{tab:1}). Spread
of experimental points leads to the statistically unacceptable
values of $\chi^{2}/\mbox{n.d.f.}$ In the case of $\lambda$
inclusion of estimation for systematic uncertainty for
$\mbox{Pb+Pb}$ collisions at
$\sqrt{\smash[b]{s_{\footnotesize{NN}}}}=2.76$ TeV leads to both
the dramatic growth of $a_{2}$ and improvement of the fit quality
at transition from statistical errors of experimental points to
total errors in data sample (i) (Table\,\ref{tab:1}). Inclusion of
total errors allows to get a statistically acceptable fit
qualities for HBT radii for both the function (\ref{eq:Fit-1}) and
its specific case. It seems more complex fit function should be
used in order to describe energy dependence of HBT radii at all
available collision energies. This study is in the progress.
Taking into account the similar behavior of the energy dependence
of HBT radii (Figs.\,\ref{fig:2}b -- d) and elliptic flow $v_{2}$
\cite{Okorokov-SPMP-201-2008} at qualitative level the following
functional form can be suggested
$g(\sqrt{\smash[b]{s_{\footnotesize{NN}}}})=a_{1}+a_{2}(\sqrt{\varepsilon}-a_{3})^{a_{4}}+\sum_{i=5,6}a_{i}\varepsilon^{a_{i+1}}+a_{9}(\ln\varepsilon)^{a_{10}}$
as first approach for description of
$\mathcal{G}_{1}^{i}(\sqrt{\smash[b]{s_{\footnotesize{NN}}}})$,
$i=1-4$ in all experimentally available energy domain. Smooth
solid and dashed curves shown in Fig.\,\ref{fig:1}e are calculated
for the ratio
$R_{\mbox{\scriptsize{o}}}/R_{\mbox{\scriptsize{s}}}$ from the fit
results for $R_{\mbox{\scriptsize{s}}}$ and
$R_{\mbox{\scriptsize{o}}}$ (Table\,\ref{tab:1}). As seen these
curves agree with experimental points reasonably at
$\sqrt{\smash[b]{s_{\footnotesize{NN}}}} \geq 5$ GeV. In general
fits by the function (\ref{eq:Fit-1}) at free $a_{3}$ and fixed
$a_{3}$ show close behavior for all the main HBT parameters from
$\mathcal{G}_{1}$ with some differences at intermediate
($\sqrt{\smash[b]{s_{\footnotesize{NN}}}} \lesssim 10$ GeV) and
high ($\sqrt{\smash[b]{s_{\footnotesize{NN}}}} > 200$ GeV)
energies. These differences result in more significant discrepancy
between fit curves for
$R_{\mbox{\scriptsize{o}}}/R_{\mbox{\scriptsize{s}}}$
(Fig.\,\ref{fig:2}e) and for other parameters from the set
$\mathcal{G}_{2}$ (see below).

\begin{table*}[h!]
\caption{Values of fit parameters for approximation of the data
sample (i)} \label{tab:1}
\begin{center}
\begin{tabular}{lcccccccc}
\hline \multicolumn{1}{l}{HBT} & \multicolumn{4}{c}{Fit with
statistical errors} &
\multicolumn{4}{c}{Fit with total errors} \rule{0pt}{10pt}\\
\cline{2-9}
parameter & $a_{1}$ & $a_{2}$ & $a_{3}$ & $\chi^{2}/\mbox{n.d.f.}$ & $a_{1}$ & $a_{2}$ & $a_{3}$ & $\chi^{2}/\mbox{n.d.f.}$ \rule{0pt}{10pt}\\
\hline
$\lambda$                   & $0.36 \pm 0.02$   & $1.90 \pm 0.12$                & $-0.91 \pm 0.14$  & $534/23$  & $0.008 \pm 0.002$ & $104 \pm 21$               & $-0.291 \pm 0.011$ & $373/23$ \rule{0pt}{10pt}\\
                            & $0.570 \pm 0.004$ & $-0.021 \pm 0.001$             & $1.0$ (fixed)     & $580/19$  & $0.621 \pm 0.005$ & $-0.032 \pm 0.001$         & $1.0$ (fixed)      & $266/19$ \rule{0pt}{10pt}\\
$R_{\mbox{\scriptsize{s}}}$ & $4.77 \pm 0.02$   & $(1.3 \pm 0.8) \times 10^{-4}$ & $2.8 \pm 0.2$     & $141/22$  & $4.5 \pm 0.2$     & $(1 \pm 2) \times 10^{-3}$ & $2.3 \pm 0.9$      & $51.1/22$ \rule{0pt}{10pt}\\
                            & $4.37 \pm 0.03$   & $0.0188 \pm 0.0012$            & $1.0$ (fixed)     & $194/23$  & $3.88 \pm 0.18$   & $0.038 \pm 0.008$          & $1.0$ (fixed)      & $54.5/23$ \rule{0pt}{10pt}\\
$R_{\mbox{\scriptsize{o}}}$ & $0.59 \pm 0.06$   & $7.4 \pm 1.0$                  & $0.097 \pm 0.009$ & $392/22$  & $0.64 \pm 0.11$   & $7 \pm 2$                  & $0.03 \pm 0.03$    & $31.9/22$ \rule{0pt}{10pt}\\
                            & $5.38 \pm 0.04$   & $0.0120 \pm 0.0012$            & $1.0$ (fixed)     & $411/23$  & $5.3 \pm 0.2$     & $0.003 \pm 0.002$          & $1.0$ (fixed)      & $31.9/23$ \rule{0pt}{10pt}\\
$R_{\mbox{\scriptsize{l}}}$ & $0.110 \pm 0.015$ & $32 \pm 4$                     & $0.267 \pm 0.008$ & $357/21$  & $0.03 \pm 0.03$   & $87 \pm 64$                & $0.33 \pm 0.03$    & $23.3/21$ \rule{0pt}{10pt}\\
                            & $4.51 \pm 0.04$   & $0.0475 \pm 0.0018$            & $1.0$ (fixed)     & $459/22$  & $4.0 \pm 0.2$     & $0.065 \pm 0.010$          & $1.0$ (fixed)      & $26.3/22$ \rule{0pt}{10pt}\\
\hline
\end{tabular}
\end{center}
\end{table*}

Fig.\,\ref{fig:2} demonstrates the energy dependence of
$\Delta\tau$ for (quasi)symmetric heavy ion collisions. The
emission duration in these collisions is calculated based on known
HBT-radii (Figs.\,\ref{fig:1}b -- d), kinematic regime for pion
pairs and on (\ref{eq:2.6dop}). The $\langle \beta_{\perp}\rangle
\approx 0.82$ for pion pairs with $\langle k_{\perp}\rangle \simeq
0.2$ GeV/$c$. Value $\Delta\tau= (0.53 \pm 9.15)$ fm/$c$ at
$\sqrt{\smash[b]{s_{\footnotesize{NN}}}}=130$ GeV derived from the
PHENIX results at this energy is not shown due to extremely large
errors. As seen the emission duration for pions extracted from
$\delta$ (\ref{eq:2.5dop}) is about $2-4$ fm/$c$ for any energies
under consideration. The visible energy dependence of emission
duration is absent,
$\Delta\tau(\sqrt{\smash[b]{s_{\footnotesize{NN}}}})$ is close to
flat within large error bars. One can see more interesting
behavior for this dependence for the STAR high-statistics data
\cite{STAR-arXiv-1403.4972} only. But additional precise
measurements are necessary in order to confirm the change of
$\Delta\tau(\sqrt{\smash[b]{s_{\footnotesize{NN}}}})$ at
$\sqrt{\smash[b]{s_{\footnotesize{NN}}}} \sim 10-20$ GeV and
locate the possible knee in the experimental dependence. Smooth
solid and dashed curves shown in Fig.\,\ref{fig:2} are calculated
for $\Delta\tau$ from the fit results for
$R_{\mbox{\scriptsize{s}}}$ and $R_{\mbox{\scriptsize{o}}}$
(Table\,\ref{tab:1}). It seems the function (\ref{eq:Fit-1}) at
free $a_{3}$ agrees better with experimental points at
$\sqrt{\smash[b]{s_{\footnotesize{NN}}}} \leq 200$ GeV than that
at fixed $a_{3}$. But large error bars do not allow to choice
preferable curve unambiguously. Moreover the general function
(\ref{eq:Fit-1}) underestimates $\Delta\tau$ in TeV-region
significantly.

\begin{figure*}
\includegraphics[width=15.5cm,height=17.0cm]{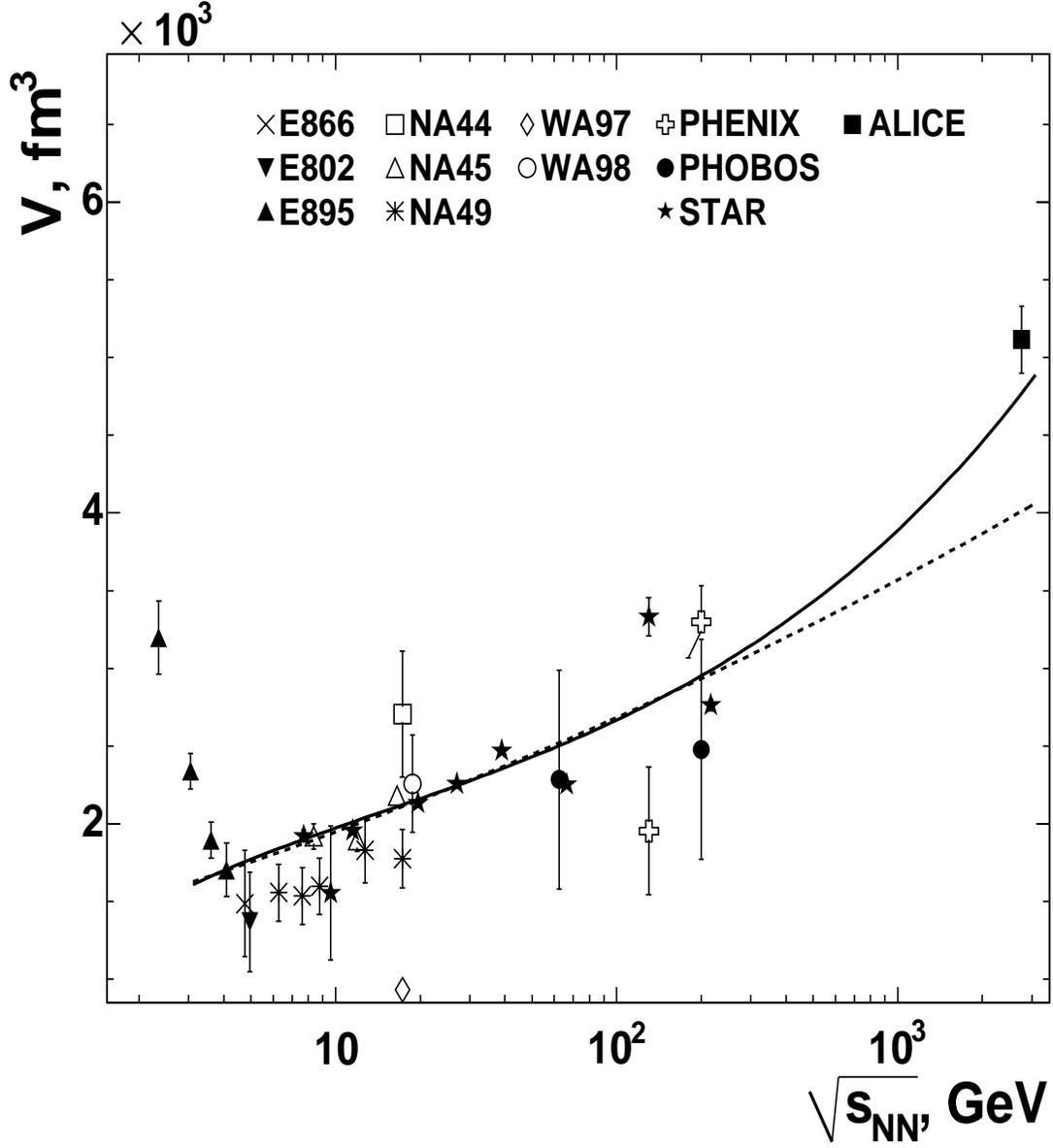}
\vspace*{8pt} \caption{Energy dependence of volume of emission
region at freeze-out for secondary charged pions in central heavy
ion collisions $\mbox{Au+Au, Au+Pb, Pb+Pb}$ in midrapidity region
and at $\langle k_{\perp}\rangle \simeq 0.2$ GeV/$c$. The equation
(\ref{eq:2.4new}) is used for calculation of volume values.
Experimental results are shown for the same particle types and
Coulomb corrections as well as in Fig.\,\ref{fig:1}. Error bars
are only statistical (for NA44 -- total uncertainties). Smooth
curves are derived from (\ref{eq:2.4new}) and the fit results for
$R_{\mbox{\scriptsize{s}}}$, $R_{\mbox{\scriptsize{l}}}$ with
taking into account the point of the WA97 experiment
\cite{Antinori-JPGNPP-27-2325-2001}. The solid line corresponds to
the fits of HBT radii by function (\ref{eq:Fit-1}) and dashed line
-- to the fits by specific case $R_{i} \propto \ln\varepsilon$,
$i=\mbox{s}, \mbox{l}$.} \label{fig:3}
\end{figure*}

Volume of the homogeneity region in various heavy ion collisions
is calculated based on known HBT-radii which are shown in
Figs.\,\ref{fig:1}b -- d. Both the equations (\ref{eq:2.4}) and
(\ref{eq:2.4new}) are used for verification and increasing of
reliability of results. The first approach is the same as in the
previous study \cite{Okorokov-arXiv-1312.4269}. As expected the
values of $V$ are similar for both the equations (\ref{eq:2.4})
and (\ref{eq:2.4new}) that confirms the validity of the results
from \cite{Okorokov-arXiv-1312.4269}. The energy dependence of
estimations (\ref{eq:2.4new}) for volume of emission region is
shown in Fig.\,\ref{fig:3}. It would be emphasized that values of
$V$ calculated at $\sqrt{\smash[b]{s_{\footnotesize{NN}}}}=62.4$
and 200 GeV from the recent STAR high-statistics results
\cite{STAR-arXiv-1403.4972} agree better, especially at top RHIC
energy, with results of other experiments (PHENIX and PHOBOS) than
that in earlier study \cite{Okorokov-arXiv-1312.4269}. The values
of $V^{(\ref{eq:2.4})}$ which are derived here from the STAR
results obtained in the framework of the phase-I beam energy
program (BES) at RHIC in energy domain
$\sqrt{\smash[b]{s_{\footnotesize{NN}}}}=7.7-39$ GeV form the
trend lies some higher than most of the results from AGS and SPS.
But on the other hand the spread of results at
$\sqrt{\smash[b]{s_{\footnotesize{NN}}}}=7-20$ GeV is within total
errors if systematic uncertainties will be taken into account
also. The values of $V^{(\ref{eq:2.4new})}$ from the STAR results
at energies $\sqrt{\smash[b]{s_{\footnotesize{NN}}}}=7.7-39$ GeV
\cite{STAR-arXiv-1403.4972} agree noticeably better with AGS and
SPS results (Fig.\,\ref{fig:3}) than that for
$V^{(\ref{eq:2.4})}$. Results for volume at
$\sqrt{\smash[b]{s_{\footnotesize{NN}}}}=7.7-39$ GeV which are
derived from the STAR high-statistics data
\cite{STAR-arXiv-1403.4972} based on (\ref{eq:2.4}) as well as on
(\ref{eq:2.4new}) agree better with (quasi)linear growth $V$ with
$\ln\varepsilon$ from $\sqrt{\smash[b]{s_{\footnotesize{NN}}}}
\simeq 5$ GeV up to highest RHIC energy than that the earlier data
at close energies. Smooth solid and dashed curves shown in
Fig.\,\ref{fig:3} are calculated for $V$ from equation
(\ref{eq:2.4new}) and the fit results for
$R_{\mbox{\scriptsize{s}}}$, $R_{\mbox{\scriptsize{l}}}$
(Table\,\ref{tab:1}). Both the curves are very close at
$\sqrt{\smash[b]{s_{\footnotesize{NN}}}} \leq 200$ GeV but
function (\ref{eq:Fit-1}) at $a_{3}=1.0$ underestimates
$V^{\ref{eq:2.4new}}$ in TeV-region significantly. Therefore the
general function (\ref{eq:Fit-1}) is the preferable approximation
of the experimental
$V^{\ref{eq:2.4new}}(\sqrt{\smash[b]{s_{\footnotesize{NN}}}})$ at
$\sqrt{\smash[b]{s_{\footnotesize{NN}}}} \geq 5$ GeV.

Predictions for values of the femtoscopic observables from sets
$\mathcal{G}_{m}$, $m=1, 2$ are obtained for heavy-ion mode
energies of the LHC and the FCC project based on the fit results
for the main HBT parameters. Estimations are shown in
Table\,\ref{tab:2} for fits with inclusion of statistical errors,
the second line for each collision energy corresponds to the using
of the specific case of (\ref{eq:Fit-1}), the volume of
homogeneity region is estimated with help of (\ref{eq:2.4new}).
Large uncertainties obtained for estimations based on the function
(\ref{eq:Fit-1}) do not allow to distinguish predictions from
(\ref{eq:Fit-1}) with free $a_{3}$ and with fixed $a_{3}=1.0$. One
can expect the volume of homogeneity region $V^{(\ref{eq:2.4new})}
\sim 6000$ fm$^{3}$ at
$\sqrt{\smash[b]{s_{\footnotesize{NN}}}}=5.52$ TeV (LHC) and
$V^{(\ref{eq:2.4new})} \sim 9000$ fm$^{3}$ at
$\sqrt{\smash[b]{s_{\footnotesize{NN}}}}=39.0$ TeV (FCC) based on
the reasonable agreement between experimental data and solid curve
at Fig.\,\ref{fig:3}.

\begin{table*}[h!]
\caption{Estimations for observables based on fit results for the
data sample (i)} \label{tab:2}
\begin{center}
\begin{tabular}{lccccccc}
\hline
\multicolumn{1}{l}{$\sqrt{\smash[b]{s_{\footnotesize{NN}}}}$,} &
\multicolumn{7}{c}{HBT parameter}\rule{0pt}{10pt}\\
\cline{2-8}
TeV & $\lambda$ & $R_{\mbox{\scriptsize{s}}}$, fm & $R_{\mbox{\scriptsize{o}}}$, fm & $R_{\mbox{\scriptsize{l}}}$, fm & $R_{\mbox{\scriptsize{o}}}/R_{\mbox{\scriptsize{s}}}$ & $\Delta\tau$, fm/$c$ & $V \times 10^{-3}$, fm$^{3}$ \rule{0pt}{10pt}\\
\hline
5.52 & $0.41 \pm 0.03$   & $6.8 \pm 1.9$   & $6.3 \pm 1.0$   & $7.6 \pm 1.5$   & $0.9 \pm 0.3$   & --            & $6 \pm 3$ \rule{0pt}{10pt}\\
     & $0.362 \pm 0.009$ & $5.79 \pm 0.10$ & $6.49 \pm 0.12$ & $8.20 \pm 0.16$ & $1.12 \pm 0.03$ & $3.6 \pm 0.4$ & $4.33 \pm 0.17$ \rule{0pt}{10pt}\\
39.0 & $0.40 \pm 0.02$   & $8 \pm 3$       & $6.4 \pm 1.0$   & $8.0 \pm 1.6$   & $0.8 \pm 0.3$   & --            & $9 \pm 8$      \rule{0pt}{10pt}\\
     & $0.315 \pm 0.011$ & $6.11 \pm 0.12$ & $6.74 \pm 0.15$ & $9.04 \pm 0.19$ & $1.10 \pm 0.03$ & $3.5 \pm 0.5$ & $5.3 \pm 0.2$ \rule{0pt}{10pt}\\
\hline
\end{tabular}
\end{center}
\end{table*}

\begin{figure*}
\includegraphics[width=15.5cm,height=17.0cm]{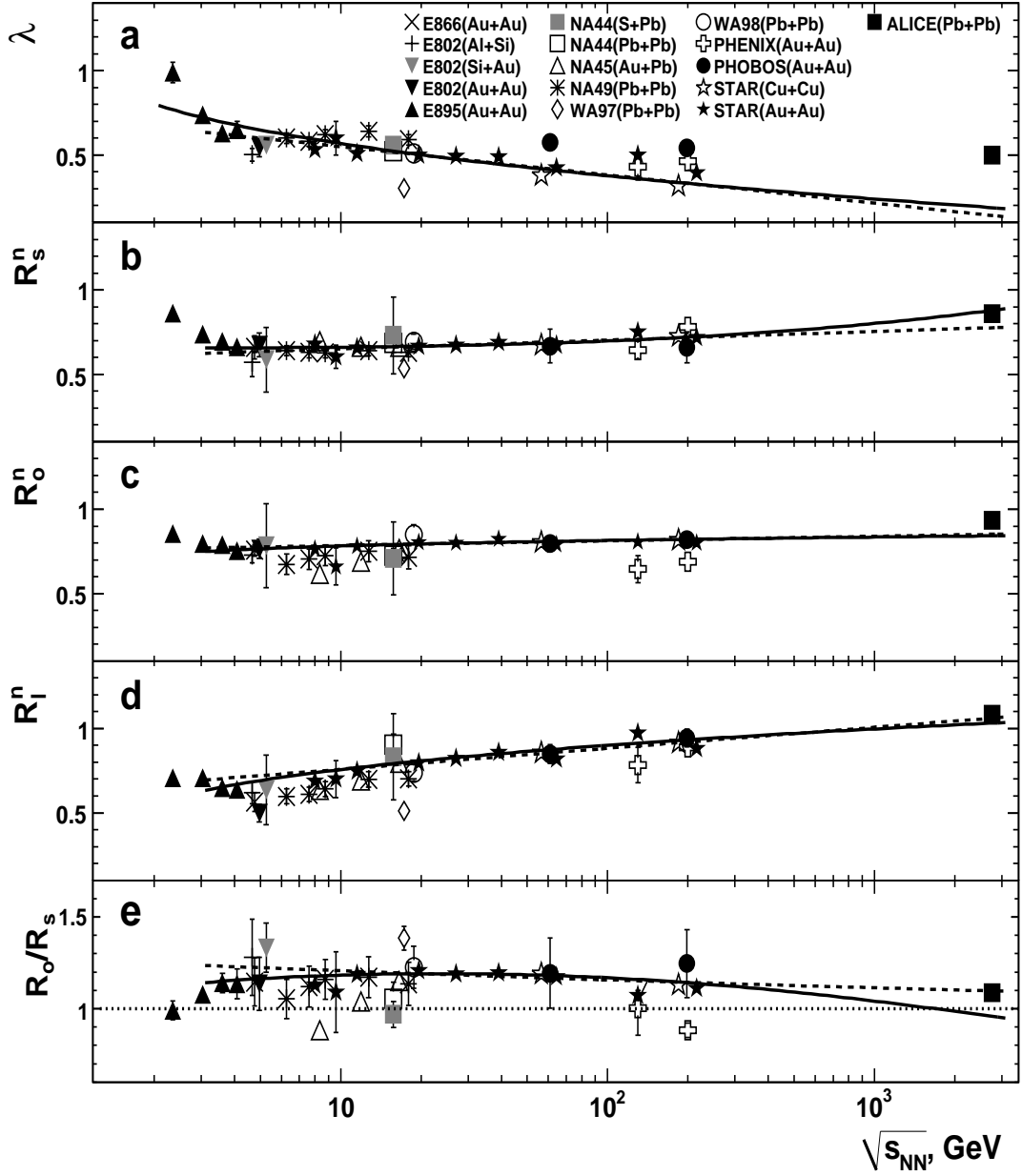}
\vspace*{8pt} \caption{Energy dependence of $\lambda$ parameter
(a), normalized HBT-radii (b -- d) and ratio
$R_{\mbox{\scriptsize{o}}}/R_{\mbox{\scriptsize{s}}}$ (e) in
various nucleus-nucleus collisions at $\langle k_{\perp}\rangle
\simeq 0.2$ GeV/$c$
\cite{Okorokov-arXiv-1312.4269,STAR-arXiv-1403.4972}. Experimental
results are shown for central collisions (for minimum bias event
in the case of E802 for $\mbox{Al+Si}$), for pairs of $\pi^{-}$
mesons (in the cases of ALICE and STAR for both the $\mbox{Cu+Cu}$
and $\mbox{Au+Au}$ at
$\sqrt{\smash[b]{s_{\footnotesize{NN}}}}=7.7-62.4$ and 200 GeV --
for $\pi^{\pm}\pi^{\pm}$ pairs, E802 for $\mbox{Al+Si}$, NA44 for
$\mbox{S+Pb}$ -- for pairs of $\pi^{+}$ mesons) and for standard
Coulomb correction $P_{\mbox{\scriptsize{C}}}^{(1)}(q)$ (in the
cases of ALICE, NA44, NA45, PHOBOS, STAR for both the
$\mbox{Cu+Cu}$ and $\mbox{Au+Au}$ at
$\sqrt{\smash[b]{s_{\footnotesize{NN}}}}=7.7, 11.5-62.4$ and 200
GeV -- for correction $P_{\mbox{\scriptsize{C}}}^{(3)}$).
Statistical errors are shown (for NA44 -- total uncertainties).
The solid lines (a -- d) correspond to the fits by function
(\ref{eq:Fit-1}) and dashed lines -- to the fits by specific case
of (\ref{eq:Fit-1}) at fixed $a_{3}=1.0$. Smooth solid and dashed
curves at (e) correspond to the ratio
$R_{\mbox{\scriptsize{o}}}/R_{\mbox{\scriptsize{s}}}$ calculated
from the fit results for $R^{n}_{\mbox{\scriptsize{s}}}$ and
$R^{n}_{\mbox{\scriptsize{o}}}$, dotted line is the level
$R_{\mbox{\scriptsize{o}}}/R_{\mbox{\scriptsize{s}}}=1$.}
\label{fig:4}
\end{figure*}

Fig.\,\ref{fig:4} shows the energy dependence of $\lambda$ (a),
scaled HBT-radii (b -- d) and
$R_{\mbox{\scriptsize{o}}}/R_{\mbox{\scriptsize{s}}}$ ratio (e)
for both the symmetric and non-symmetric collisions of various
nuclei. Fits of experimental dependencies for the data sample (ii)
are made by (\ref{eq:Fit-1}) in the same energy domains and with
the same error types as well as for the data sample (i) above. It
seems the $\lambda$ value from the WA97 experiment
\cite{Antinori-JPGNPP-27-2325-2001} can not be excluded from the
data sample (ii) because there are the STAR results $\lambda \sim
0.3$ for $\mbox{Cu+Cu}$ collisions also (Fig.\,\ref{fig:4}a).
There are no physics reasons in order to exclude the points of
these experiments from the fitted data sample (ii) in the case of
all available nucleus-nucleus collisions. Furthermore the scaled
value of longitudinal radius $R^{n}_{\mbox{\scriptsize{l}}}$ from
\cite{Antinori-JPGNPP-27-2325-2001} agrees better with results of
other experiments at close energies (Fig.\,\ref{fig:4}d) than that
for the data sample (i). Therefore there is no exception of any
experimental point from fitted ensemble for any HBT observable in
Fig.\,\ref{fig:4} in contrast with the fitting procedure for the
data sample (i). The numerical values of fit parameters are
presented in Table\,\ref{tab:3}, where the second line for
$\lambda$ and each normalized HBT radius corresponds to the
approximation by specific case of (\ref{eq:Fit-1}). Fit curves are
shown in Fig.\,\ref{fig:4} by solid lines for (\ref{eq:Fit-1}) and
by dashed lines for specific case of fit function at $a_{3}=1.0$
with taking into account statistical errors. Fit qualities are
improved for $R^{n}_{\mbox{\scriptsize{s}}}$ in the case of total
errors of experimental point and for
$R^{n}_{\mbox{\scriptsize{o}}}$ at any error types of experimental
point with respect to the corresponding fit results for the data
sample (i) shown in Table\,\ref{tab:1}. There is dramatic growth
of $\chi^{2}/\mbox{n.d.f.}$ values for fits of $\lambda$ data
(Fig.\,\ref{fig:4}a) despite of qualitative agreement between
smooth approximations and experimental $\lambda$ values for range
$10 \lesssim \sqrt{\smash[b]{s_{\footnotesize{NN}}}} \lesssim 200$
GeV. The fit by (\ref{eq:Fit-1}) at $a_{3}=1.0$ underestimates the
$\lambda$ value at the LHC energy
$\sqrt{\smash[b]{s_{\footnotesize{NN}}}}=2.76$ TeV significantly.
The $\lambda$ values for asymmetric nucleus-nucleus collisions at
intermediate energies $\sqrt{\smash[b]{s_{\footnotesize{NN}}}}
\lesssim 20$ GeV agree well with values of $\lambda$ in symmetric
heavy ion collisions at close energies. On the other hand the
$\lambda$ for $\mbox{Cu+Cu}$ collisions is smaller systematically
than $\lambda$ in $\mbox{Au+Au}$ collisions in energy range
$\sqrt{\smash[b]{s_{\footnotesize{NN}}}}=62-200$ GeV
(Fig.\,\ref{fig:4}a). New experimental data are important for
verification of the suggestion of separate dependencies
$\lambda(\sqrt{\smash[b]{s_{\footnotesize{NN}}}})$ for moderate
and heavy ion collisions. Also the development of some approach is
required in order to account for type of colliding beams in the
case of $\lambda$ parameter and improve quality of smooth
approximation. In this case significant growth of $a_{2}$ as well
as improvement of the fit quality at transition from statistical
errors of experimental points to total errors in the data sample
(ii) (Table\,\ref{tab:3}) is dominated by inclusion of estimations
for systematic uncertainties for $\mbox{Cu+Cu}$ collisions
and\,/\,or $\mbox{Pb+Pb}$ ones at
$\sqrt{\smash[b]{s_{\footnotesize{NN}}}}=2.76$ TeV. Smooth curves
for normalized HBT radii and ratio
$R_{\mbox{\scriptsize{o}}}/R_{\mbox{\scriptsize{s}}}$ are in
reasonable agreement with experimental dependencies in fitted
domain of collision energies
$\sqrt{\smash[b]{s_{\footnotesize{NN}}}} \geq 5$ GeV
(Figs.\,\ref{fig:4}b -- e). Parameter values obtained for fit of
$R^{n}_{\mbox{\scriptsize{l}}}$ with total uncertainties by
(\ref{eq:Fit-1}) at $a_{3}=1.0$ are equal within errors with
results from \cite{Alexander-JPGNPP-40-125101-2013} accounting for
that experimental results studied here are obtained at $\langle
m_{\perp}\rangle \simeq 1.75m_{\pi}$. Dramatic improvement of the
fit qualities for scaled HBT radii at transition from the data
sample (ii) with statistical errors to the data sample with total
errors (Table\,\ref{tab:3}) is dominated mostly by the uncertainty
in $r_{0}$ leads to additional errors due to scaling
(\ref{eq:2.8}). At the same time inclusion of total uncertainties
for $\mbox{Au+Au}$ collisions at
$\sqrt{\smash[b]{s_{\footnotesize{NN}}}}=19.6$ GeV results in
significant decreasing of $a_{2}$ parameter in the case of
$R^{n}_{\mbox{\scriptsize{l}}}$ scaled radius.

\begin{table*}[h!]
\caption{Values of fit parameters for approximation of the data
sample (ii)} \label{tab:3}
\begin{center}
\begin{tabular}{lcccccccc}
\hline \multicolumn{1}{l}{HBT} & \multicolumn{4}{c}{Fit with
statistical errors} &
\multicolumn{4}{c}{Fit with total errors} \rule{0pt}{10pt}\\
\cline{2-9}
parameter & $a_{1}$ & $a_{2}$ & $a_{3}$ & $\chi^{2}/\mbox{n.d.f.}$ & $a_{1}$ & $a_{2}$ & $a_{3}$ & $\chi^{2}/\mbox{n.d.f.}$ \rule{0pt}{10pt}\\
\hline
$\lambda$                       & $1.21 \pm 0.09$   & $-0.30 \pm 0.04$           & $0.38 \pm 0.04$   & $3656/29$ & $0.60 \pm 0.02$   & $-0.014 \pm 0.008$         & $1.3 \pm 0.2$   & $780/29$ \rule{0pt}{10pt}\\
                                & $0.717 \pm 0.003$ & $-0.051 \pm 0.001$         & $1.0$ (fixed)     & $3786/23$ & $0.631 \pm 0.005$ & $-0.034 \pm 0.001$         & $1.0$ (fixed)   & $706/23$ \rule{0pt}{10pt}\\
$R^{n}_{\mbox{\scriptsize{s}}}$ & $0.656 \pm 0.002$ & $(6 \pm 3) \times 10^{-5}$ & $3.11 \pm 0.19$   & $195/25$  & $0.63 \pm 0.02$   & $(6 \pm 5) \times 10^{-4}$ & $2.4 \pm 0.9$   & $26.8/25$ \rule{0pt}{10pt}\\
                                & $0.599 \pm 0.003$ & $0.019 \pm 0.001$          & $1.0$ (fixed)     & $280/26$  & $0.56 \pm 0.03$   & $0.029 \pm 0.008$          & $1.0$ (fixed)   & $28.9/26$ \rule{0pt}{10pt}\\
$R^{n}_{\mbox{\scriptsize{o}}}$ & $0.10 \pm 0.02$   & $6.3 \pm 1.7$              & $0.068 \pm 0.006$ & $402/25$  & $0.019 \pm 0.003$ & $30 \pm 9$                 & $0.12 \pm 0.05$ & $23.9/25$ \rule{0pt}{10pt}\\
                                & $0.758 \pm 0.004$ & $0.008 \pm 0.001$          & $1.0$ (fixed)     & $415/26$  & $0.67 \pm 0.04$   & $0.017 \pm 0.008$          & $1.0$ (fixed)   & $24.4/26$ \rule{0pt}{10pt}\\
$R^{n}_{\mbox{\scriptsize{l}}}$ & $0.022 \pm 0.002$ & $23 \pm 3$                 & $0.258 \pm 0.005$ & $502/25$  & $0.23 \pm 0.04$   & $0.8 \pm 0.2$              & $0.57 \pm 0.05$ & $66.0/25$ \rule{0pt}{10pt}\\
                                & $0.634 \pm 0.004$ & $0.043 \pm 0.001$          & $1.0$ (fixed)     & $615/26$  & $0.47 \pm 0.03$   & $0.089 \pm 0.014$          & $1.0$ (fixed)   & $66.7/26$ \rule{0pt}{10pt}\\
\hline
\end{tabular}
\end{center}
\end{table*}

\begin{figure*}
\includegraphics[width=15.5cm,height=17.0cm]{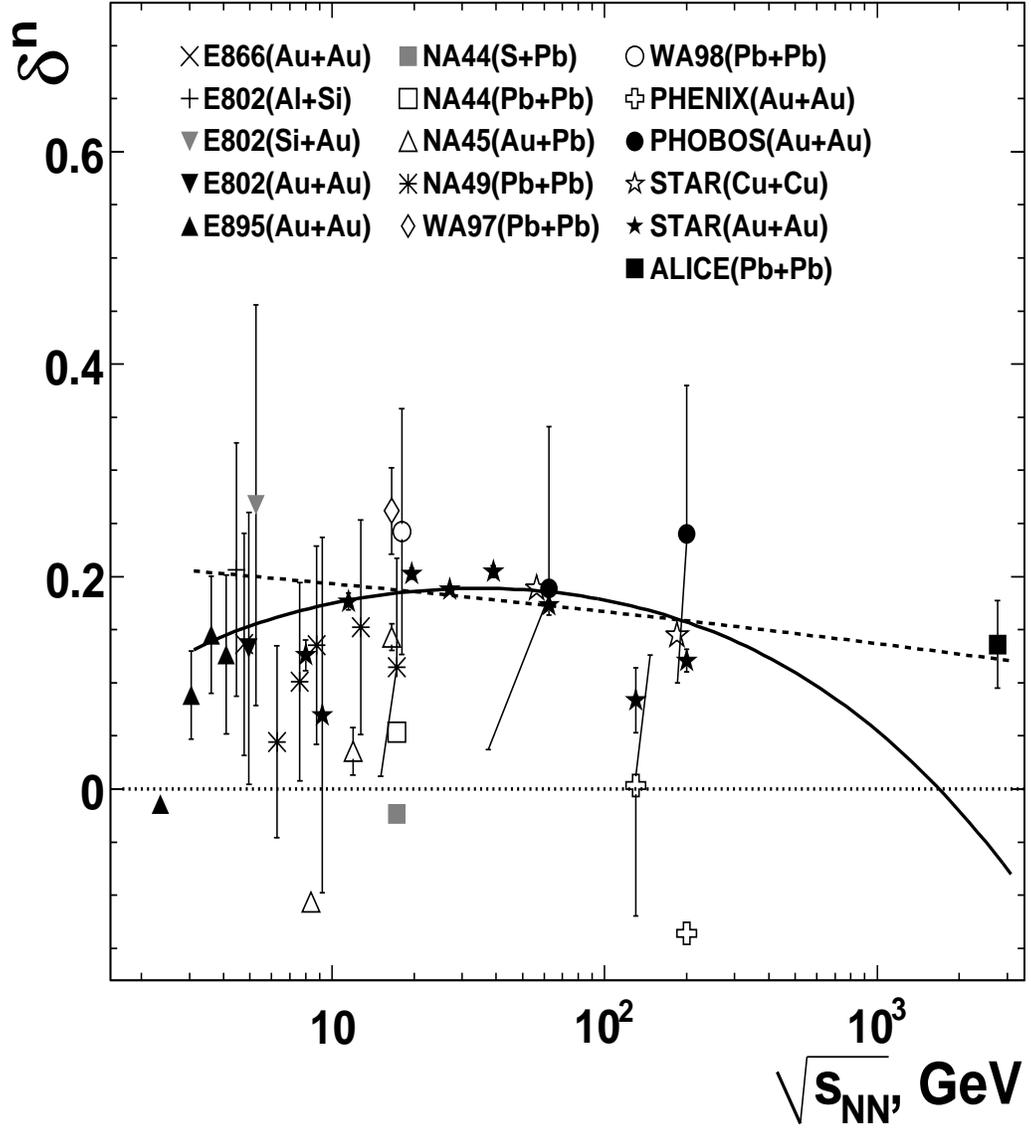}
\vspace*{8pt} \caption{Dependence of scaled difference of squares
of transverse radii on beam energy for emission region of
secondary charged pion in various nucleus-nucleus collisions at
$\langle k_{\perp}\rangle \simeq 0.2$ GeV/$c$. Experimental
results are shown for the same particle types and Coulomb
corrections as well as in Fig.\,\ref{fig:4}. Error bars are only
statistical (for NA44 -- total uncertainties). Dotted line is the
level $\delta^{n}=0$. Smooth curves are derived from
(\ref{eq:2.8}) and the fit results for
$R^{n}_{\mbox{\scriptsize{s}}}$, $R^{n}_{\mbox{\scriptsize{o}}}$.
The solid line corresponds to the fits of normalized HBT radii by
function (\ref{eq:Fit-1}) and dashed line -- to the fits by
specific case $R^{n}_{i} \propto \ln\varepsilon$, $i=\mbox{s},
\mbox{o}$.} \label{fig:5}
\end{figure*}

\begin{figure*}
\includegraphics[width=15.5cm,height=17.0cm]{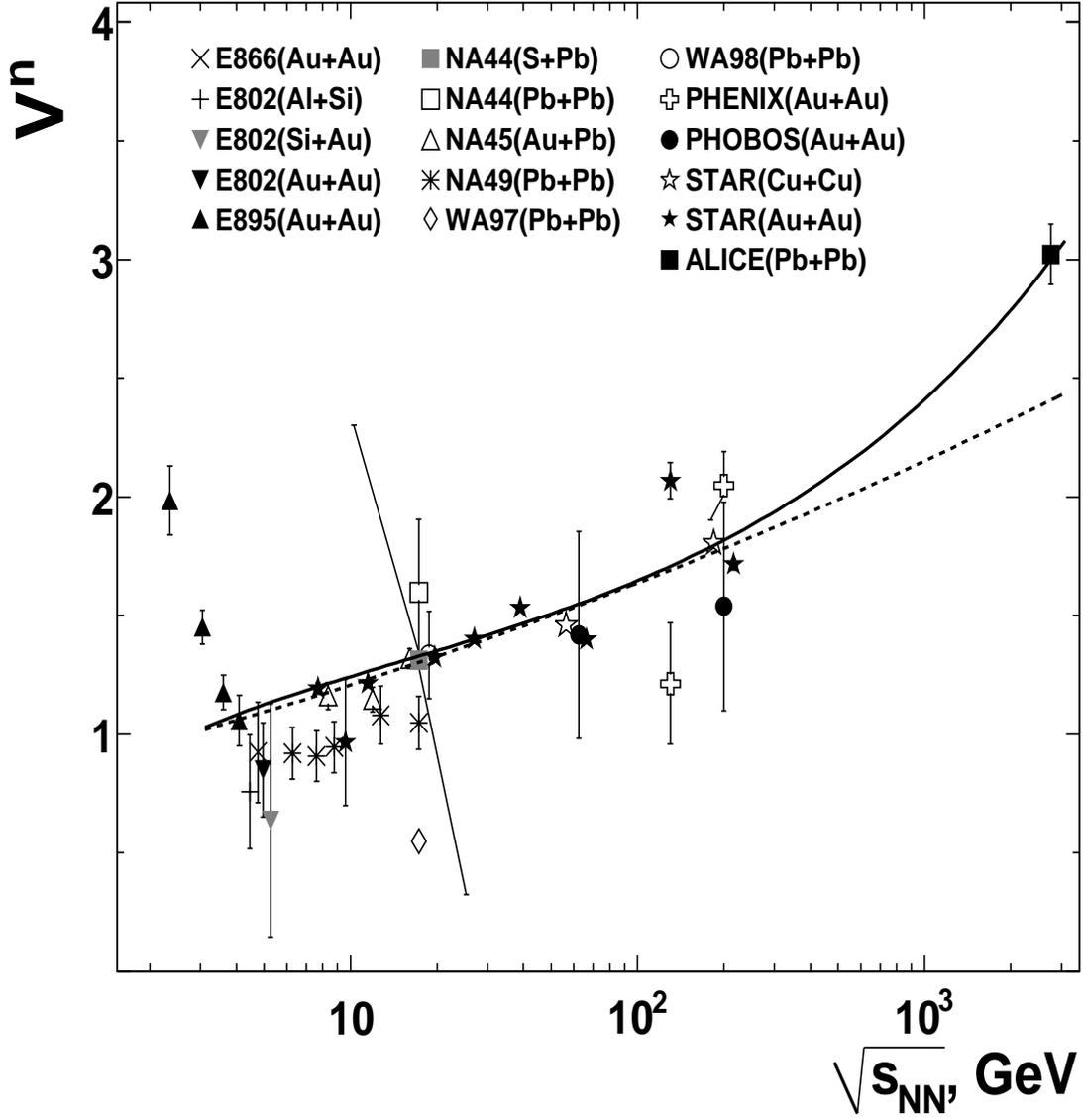}
\vspace*{8pt} \caption{Energy dependence of normalized volume of
emission region at freeze-out for secondary charged pions in
various nucleus-nucleus collisions at $\langle k_{\perp}\rangle
\simeq 0.2$ GeV/$c$. The equation (\ref{eq:2.4new}) is used for
calculation of volume values. Experimental results are shown for
the same collision, particle and Coulomb correction types as well
as in Fig.\,\ref{fig:4}. Error bars are only statistical (for NA44
-- total uncertainties). Smooth curves are derived from
(\ref{eq:2.4new}) and the fit results for
$R^{n}_{\mbox{\scriptsize{s}}}$, $R^{n}_{\mbox{\scriptsize{l}}}$.
The solid line corresponds to the fits of normalized HBT radii by
function (\ref{eq:Fit-1}) and dashed line -- to the fits by
specific case $R^{n}_{i} \propto \ln\varepsilon$, $i=\mbox{s},
\mbox{l}$.} \label{fig:6}
\end{figure*}

The corresponding dependencies for $\delta^{n}$ and $V^{n}$ are
demonstrated in Fig.\,\ref{fig:5} and Fig.\,\ref{fig:6},
respectively. The definition (\ref{eq:2.4new}) for source volume
is used in Fig.\,\ref{fig:6} as well as in Fig.\,\ref{fig:3}
above. As well as in \cite{Okorokov-arXiv-1312.4269} results for
$\pi^{+}\pi^{+}$ pairs are shown in Figs.\,\ref{fig:4} --
\ref{fig:6} also because fem\-to\-sco\-py parameters from the set
$\mathcal{G}_{1}$ depend on sign of electrical charge of secondary
pions weakly. The relation $R_{\mbox{\scriptsize{o}}} <
R_{\mbox{\scriptsize{s}}}$ is observed for $\approx 11\%$ of
points in Fig.\,\ref{fig:5}. In general the $\delta < 0$ can be
possible in the model of opaque source with surface dominated
emission
\cite{Heiselberg-EPJC-1-593-1998,McLerran-arXiv-nucl-th-0205028}.
But possibly results should be similar for both the same ion beams
and close kinematic regimes in various experiments. Therefore
additional study is required in order to distinguish the physical
and technique sources of negative values of the $\delta^{n}$ in
Fig.\,\ref{fig:5} and to get a more definite explanation. The
dependence $\delta^{n}(\sqrt{\smash[b]{s_{\footnotesize{NN}}}})$
is almost flat within large error bars in all energy domain under
consideration. Taking into account the STAR high-statistics
results \cite{STAR-arXiv-1403.4972} only one can see the
indication on change of behavior of
$\delta^{n}(\sqrt{\smash[b]{s_{\footnotesize{NN}}}})$ inside the
range of collision energy
$\sqrt{\smash[b]{s_{\footnotesize{NN}}}}=11.5-19.6$ GeV. This
observation is in agreement with features of behavior of emission
duration dependence on $\sqrt{\smash[b]{s_{\footnotesize{NN}}}}$
(Fig.\,\ref{fig:2}) discussed above. The estimation of energy
range agrees well with results of several studies
\cite{STAR-arXiv-1403.4972,STAR-PRC-88-014902-2013,STAR-PRL-110-142301-2013,STAR-PRL-112-032302-2014,
STAR-PRL-112-162301-2014,STAR-PRL-113-092301-2014,STAR-PRL-113-052302-2014}
in the framework of the phase-I of BES program at RHIC which
indicate on the transition from dominance of quark-gluon degrees
of freedom to hadronic matter at
$\sqrt{\smash[b]{s_{\footnotesize{NN}}}} \lesssim 19.6$ GeV. But
future precise measurements are crucially important for extraction
of more definite physics conclusions. Smooth solid and dashed
curves shown in Fig.\,\ref{fig:5} are calculated for $\delta^{n}$
from the fit results for $R^{n}_{\mbox{\scriptsize{s}}}$ and
$R^{n}_{\mbox{\scriptsize{o}}}$ (Table\,\ref{tab:3}). The
situation is similar to that for $\Delta\tau$: calculation based
on the fit function (\ref{eq:Fit-1}) at free $a_{3}$ agrees
reasonably with experimental points at
$\sqrt{\smash[b]{s_{\footnotesize{NN}}}} \leq 200$ GeV but
underestimates $\delta^{n}$ in TeV-region significantly. The large
errors in Fig.\,\ref{fig:6} for strongly asymmetric nuclear
collisions is dominated by large difference of radii of colliding
moderate and heavy nuclei and corresponding large uncertainty for
$\langle R_{\mbox{\scriptsize{A}}}\rangle$. Smooth solid and
dashed curves shown in Fig.\,\ref{fig:6} are calculated for
$V^{n}$ from equation (\ref{eq:2.4new}) and the fit results for
$R^{n}_{\mbox{\scriptsize{s}}}$, $R^{n}_{\mbox{\scriptsize{l}}}$
(Table\,\ref{tab:3}). The fit results for normalized HBT radii
obtained with general function (\ref{eq:Fit-1}) lead to very good
agreement between smooth curve and experimental data in TeV-region
in contrast with the curve obtained from corresponding fit results
for (\ref{eq:Fit-1}) at $a_{3}=1.0$.

Estimations for $\lambda$,
$R_{\mbox{\scriptsize{o}}}/R_{\mbox{\scriptsize{s}}}$, and
normalized femtoscopic parameters at the LHC and the FCC energies
are shown in Table\,\ref{tab:4} for fits of various
nucleus-nucleus collisions with inclusion of statistical errors,
the second line for each collision energy corresponds to the using
of the specific case of (\ref{eq:Fit-1}) at $a_{3}=1.0$, the
volume of homogeneity region is estimated with help of
(\ref{eq:2.4new}). All the smooth approximations discussed above
predict amplification of coherent pion emission with significant
decreasing of $\lambda$. Uncertainties are large for estimations
obtained on the basis of results of fits by function
(\ref{eq:Fit-1}) at free $a_{3}$. Thus values of femtoscopic
observables in Table\,\ref{tab:4} are equal within errors for
general and specific case of (\ref{eq:Fit-1}) at
$\sqrt{\smash[b]{s_{\footnotesize{NN}}}}=5.52$ TeV (LHC) and
$\sqrt{\smash[b]{s_{\footnotesize{NN}}}}=39.0$ TeV (FCC) as well
as for estimations obtained on basis of the data sample (i) above.

\begin{table*}[h!]
\caption{Estimations for observables based on fit results for the
data sample (ii)} \label{tab:4}
\begin{center}
\begin{tabular}{lccccccc}
\hline
\multicolumn{1}{l}{$\sqrt{\smash[b]{s_{\footnotesize{NN}}}}$,} &
\multicolumn{7}{c}{HBT parameter}\rule{0pt}{10pt}\\
\cline{2-8}
TeV & $\lambda$ & $R^{n}_{\mbox{\scriptsize{s}}}$ & $R^{n}_{\mbox{\scriptsize{o}}}$ & $R^{n}_{\mbox{\scriptsize{l}}}$ & $R_{\mbox{\scriptsize{o}}}/R_{\mbox{\scriptsize{s}}}$ & $\delta^{n}$ & $V^{n}$ \rule{0pt}{10pt}\\
\hline
5.52 & $0.16 \pm 0.19$   & $0.9 \pm 0.2$     & $0.8 \pm 0.3$     & $1.06 \pm 0.16$   & $0.9 \pm 0.4$     & $-0.2 \pm 0.6$  & $3.5 \pm 1.6$ \rule{0pt}{10pt}\\
     & $0.091 \pm 0.004$ & $0.792 \pm 0.009$ & $0.860 \pm 0.010$ & $1.099 \pm 0.013$ & $1.086 \pm 0.018$ & $0.11 \pm 0.02$ & $2.59 \pm 0.07$ \rule{0pt}{10pt}\\
39.0 & $0.07 \pm 0.21$   & $1.2 \pm 0.4$     & $0.9 \pm 0.3$     & $1.11 \pm 0.16$   & $0.7 \pm 0.3$     & $-0.7 \pm 1.1$  & $6 \pm 4$      \rule{0pt}{10pt}\\
     & --                & $0.836 \pm 0.011$ & $0.883 \pm 0.012$ & $1.205 \pm 0.015$ & $1.06 \pm 0.02$   & $0.08 \pm 0.03$ & $3.17 \pm 0.09$ \rule{0pt}{10pt}\\
\hline
\end{tabular}
\end{center}
\end{table*}

The energy dependencies for sets $\mathcal{G}_{m}$, $m=1, 2$ of
femtoscopic parameters with taking into account the scaling
relation (\ref{eq:2.8}) and the high-statistics STAR data
\cite{STAR-arXiv-1403.4972} demonstrate the reasonable agreement
between values of parameters obtained for interactions of various
ions (shown in Figs.\,\ref{fig:4} -- \ref{fig:6}). The observation
confirms the suggestion \cite{Okorokov-arXiv-1312.4269} that
normalized femtoscopic parameters allow to unite the study both
the symmetric and asymmetric nucleus-nucleus collisions in the
framework of united approach. This qualitative suggestion is
confirmed indirectly by recent study of two-pion correlations in
the collisions of the lightest nucleus ($\mbox{d}$) with heavy ion
($\mbox{Au}$) at RHIC. Estimations of space-time extent of the
pion emission source in $\mbox{d+Au}$ collisions at top RHIC
energy \cite{PHENIX-arXiv-1404.5291} in dependence on kinematic
observables (collision centrality, the mean transverse momentum
for pion pairs) indicate similar patterns with corresponding
dependencies in $\mbox{Au+Au}$ collisions and indicate on
similarity in expansion dynamics in collisions of various systems
($\mbox{d+Au}$ and $\mbox{Au+Au}$ at RHIC, $\mbox{p+Pb}$ and
$\mbox{Pb+Pb}$ at LHC). The scaling results for some radii
indicate that hydrodynamic-like collective expansion is driven by
final-state rescattering effects \cite{PHENIX-arXiv-1404.5291}. On
the other hand the normalized femtoscopic parameters allow to get
the common kinematic dependencies only without any additional
information about possible general dynamic features in different
collisions. Thus the hypothesis discussed above is qualitative
only. The future quantitative theoretical and phenomenological
studies are essential for verification of general features of soft
stage dynamics for different collisions at high energies.

\section{Summary}\label{sec:4}
The following conclusions can be obtained by summarizing of the
basic results of the present study.

Energy dependence is investigated for range of all experimentally
available initial energies and for estimations of the main
femtoscopic parameters from set the $\mathcal{G}_{1}$ ($\lambda$
and radii) derived in the framework of Gauss approach as well as
for the set of important additional observables $\mathcal{G}_{2}$
contains ratio of transverse radii, emission duration (or
$\delta$) and HBT volume. There is no dramatic change of
femtoscopic parameter values with increasing of
$\sqrt{\smash[b]{s_{\footnotesize{NN}}}}$ in domain of collision
energies $\sqrt{\smash[b]{s_{\footnotesize{NN}}}} \geq 5$ GeV. The
estimation of emission duration of pions is about $2-4$ fm/$c$ for
any energies under consideration. The energy dependence is almost
flat for both the emission duration and the $\delta^{n}$ parameter
within large error bars. The indication on possible curve knee at
$\sqrt{\smash[b]{s_{\footnotesize{NN}}}} \sim 10-20$ GeV obtained
in the STAR high-statistics data agree with other results in the
framework of the phase-I of the beam energy scan program at RHIC.
But additional precise measurements are crucially important at
various $\sqrt{\smash[b]{s_{\footnotesize{NN}}}}$ in order to
confirm this feature in energy dependence of additional
femtoscopic parameters
($R_{\mbox{\scriptsize{o}}}/R_{\mbox{\scriptsize{s}}}, \Delta\tau,
\delta^{n}$).

Analytic function is suggested for approximation of energy
dependence of main HBT parameters. The fit curves demonstrate
qualitative agreement with experimental data for $\lambda$ at all
available collision energies and for both the absolute and
normalized HBT radii in energy domain
$\sqrt{\smash[b]{s_{\footnotesize{NN}}}} \geq 5$ GeV. Reasonable
fit qualities are obtained for HBT radii at approximation of
experimental points with total errors. Smooth curves calculated
for energy dependence of the set $\mathcal{G}_{2}$ of additional
femtoscopic parameters agree reasonably with corresponding
experimental data in the most cases. Estimations of femtoscopic
observables are obtained on the basis of the fit results for
energies of the LHC and the FCC project. For multi-TeV energy
domain the emission region of pions will be characterized by
decreased chaoticity parameter, linear sizes about $8.5 - 9.5$ fm
in longitudinal direction and $7 - 8$ fm in transverse plane,
volume of about $10^{4}$ fm$^{3}$.

\newpage
\newpage
\newpage
%
\newpage
\newpage

\end{document}